# Comparison between Time Shifting Deviation and Cross-correlation Methods

Zhongwang Pang, Guan Wang, Bo Wang, *Senior Member, IEEE*, and Lijun Wang, *Fellow, OSA*

*Abstract*—Time delay estimation (TDE) is an important step to identify and locate vibration source. The TDE result can be obtained by cross-correlation method through seeking the maximum correlation peak of two signals. However, the cross-correlation method will induce random error when dealing with the nonstationary signal. We propose a novel time shifting deviation (TSDEV) method to solve this problem, which has been proved to achieve ultrahigh precision localization result in the fiber vibration monitoring system. This paper compares TSDEV method with cross-correlation in detail by simulating TDE process in different conditions, such as signals with arbitrary intercepted length, nonstationary drift and correlated noise. Besides, experimental demonstration has been carried out on 60 km fiber to localize a wide band vibration signal. The typical localization error is 2 m with standard deviation of 21.4 m using TSDEV method. It stands in clear contrast to the result of cross-correlation method, whose localization error is 70 m and the standard deviation is 208.4 m. Compared with cross-correlation method, TSDEV has the same resistance to white noise, but has fewer boundary conditions and better suppression on linear drift or common noise, which leads to more precise TDE results.

## I. INTRODUCTION

TIME delay estimation (TDE) is a key step in many application scenarios such as navigation [1-2], power delivery [3-4], medical imaging [5], vibration monitoring [6-12], and acoustic sensing [13-19] etc. In practice, cross-correlation method is always used to estimate the time delay value through seeking the maximum correlation peak of two signals. It has advantages of simple operation and high estimation accuracy when dealing with the stationary random signal [19-22]. While, as pointed out in [23, 24], the applicability of cross-correlation method will be questioned when dealing with the real-world nonstationary signal, for example, the vibration induced phase changing signal detected by laser interferometry [11-12].

In these fiber vibration sensing systems, vibration event is usually localized via time delay estimation between two counter-propagating signals. The accuracy of TDE directly determines that of localization result. However, in the process of further improving the localization accuracy, we find that cross-correlation method will induce obvious error. Firstly, cross-correlation method has requirement on the length of signal [25], but the measurement time and intercepted length of signal are always limited or random when processing wideband signal. Secondly, the linear drift and low-frequency noise caused by the measurement system and environment will worsen the localization accuracy of vibration event. Recently, a structure with frequency-shifted optical delay line has been used to reduce the adverse effect of drift, while it is still hard to localize the low frequency vibrations [26]. Beyond that, we find that common wideband noise will make the TDE result shift to the direction of zero time delay. To solve these problems of cross-correlation method, we propose the time shifting deviation (TSDEV) method, which has been demonstrated capable to realize ultrahigh precision localizing during fiber vibration monitoring [27].

In this paper, we compare cross-correlation and TSDEV methods in detail to demonstrate the advantage of TSDEV method. Several situations, including arbitrary intercepted length, nonstationary drift and common noise, are simulated and analyzed. In addition, experimental demonstration has been carried out on 60 km fiber to localize a wideband low signal-to-noise ratio (SNR) vibration signal. The typical localization error is 2 m with standard deviation of 21.4 m for the TSDEV method. It stands in clear contrast to the error of 70 m with 208.4 m standard deviation for cross correlation method. Compared with cross-correlation, TSDEV method has the same resistance to white noise, but has fewer boundary conditions and better suppression effect on temperature drift or common noise.

## II. THEORY OF CROSS-CORRELATION AND TSDEV

In practical application, cross-correlation and TSDEV methods are both used to compare multiple time series and objectively determine how well they match up with each other.

This work was supported by National Natural Science Foundation of China under Grants No. 62171249, No. 61971259, and No. 91836301. (*Corresponding author: Bo Wang.*)

The authors are with the State Key Laboratory of Precision Measurement Technology and Instruments, Department of Precision Instrument, Tsinghua University, Beijing 100084, China and also with the Key Laboratory of Photonic Control Technology (Tsinghua University), Ministry of Education, Beijing, 100084, China (e-mail: pzw21@mails.tsinghua.edu.cn, g-wang19@mails.tsinghua.edu.cn, bo.wang@tsinghua.edu.cn, lwan@tsinghua.edu.cn).

In the case of fiber-based vibration sensing laser interferometry [7-12], these two methods are both used to estimate the time delay between two detected signals $x_1(t)$ and $x_2(t)$,

$$x_1(t) = s(t) + n_1(t), \quad (1)$$
$$x_2(t) = s(t-\tau_0) + n_2(t). \quad (2)$$

Here, $s(t)$ represents the vibration induced phase changing signal. $\tau_0$ is the time delay that we want to estimate. $n_1(t)$ and $n_2(t)$ are noise induced during the measurement, which are assumed to be uncorrelated with $s(t)$. Consequently, the cross-correlation function between $x_1(t)$ and $x_2(t)$ is given by:

$$R(\tau) = \lim_{T\to\infty} \frac{1}{T}\int_0^T [x_1(t)x_2(t+\tau)]dt$$
$$= \lim_{T\to\infty} \frac{1}{T}\int_0^T [s(t)s(t-\tau_0+\tau)]dt + \lim_{T\to\infty}\frac{1}{T}\int_0^T [n_1(t)n_2(t+\tau)]dt \quad (3)$$
$$= R_s(\tau) + R_n(\tau).$$

Here, $R_s(\tau)$ is the cross-correlation function between $s(t)$ and $s(t-\tau_0)$, $R_n(\tau)$ is the cross-correlation function between $n_1(t)$ and $n_2(t)$. Taking sinusoidal wave as an example, (3) can be written as:

$$R(\tau) = \frac{1}{T}\int_0^T \{[\sin(\omega t)+n_1(t)][\sin[\omega(t-\tau_0+\tau)]+n_2(t+\tau)]\}dt$$
$$= \frac{1}{2}\cos[\omega(\tau-\tau_0)] + R_n(\tau). \quad (4)$$

Here, $T$ is the period of sine wave and $\omega$ represents its frequency. $R_s(\tau) = \frac{1}{2}\cos[\omega(\tau-\tau_0)]$ will get the maximum value at $\tau = \tau_0$.

Recently, we propose a novel TSDEV method [27], and it essentially uses the square deviation of the difference between two signals as the TDE indicator. Through seeking the minimum TSDEV value, one can determine the time delay between two detected signals. The TSDEV value between $x_1(t)$ and $x_2(t)$ is shown as:

$$TSDEV^2(\tau) = \frac{1}{T}\int_0^T [x_1(t) - x_2(t+\tau) - C(\tau)]^2 dt \quad (5)$$
$$= TSDEV_s^2(\tau) + TSDEV_n^2(\tau),$$

where $C(\tau) = \frac{1}{T}\int_0^T [x_1(t) - x_2(t+\tau)]dt$ is the mean of $[x_1(t) - x_2(t+\tau)]$. $TSDEV_s^2(\tau)$ is the TSDEV value between signal $s(t)$ and $s(t-\tau_0)$, $TSDEV_n^2(\tau)$ is the TSDEV value between noise $n_1(t)$ and $n_2(t)$. Similarly, due to signal and noise are uncorrelated, the variance of sum is equal to the sum of variance ($Var(X+Y) = Var(X) + Var(Y)$)[28]. For sine wave signal, it can be written as:

$$TSDEV^2(\tau)$$
$$= \frac{1}{T}\int_0^T [\sin(\omega t) - \sin[\omega(t-\tau_0+\tau)] - C(\tau)]^2 dt + TSDEV_n^2(\tau) \quad (6)$$
$$= 1 - \cos[\omega(\tau-\tau_0)] + TSDEV_n^2(\tau).$$

In (6), $TSDEV_s^2(\tau) = 1 - \cos[\omega(\tau-\tau_0)]$ reaches the minimum value when $\tau = \tau_0$. Considering that arbitrary signal can be decomposed into signals with single frequency by Fourier series expansion (FS), this conclusion is still reliable for arbitrary signal.

In noise free cases ($n_1(t) = n_2(t) = 0$), both methods can obtain accurate TDE results when signal's length is infinite or intercepted length is integer multiples of one period. If the influence of noise is considered, there will be error on the TDE results.

## III. SIMULATION AND ANALYSIS

### A. The case of white noise

We mainly consider two kinds of noise: uncorrelated white noise ($n_w(t)$) and correlated common noise ($n_c(t)$). We firstly add uncorrelated white noise and analyze its influence on TDE results.

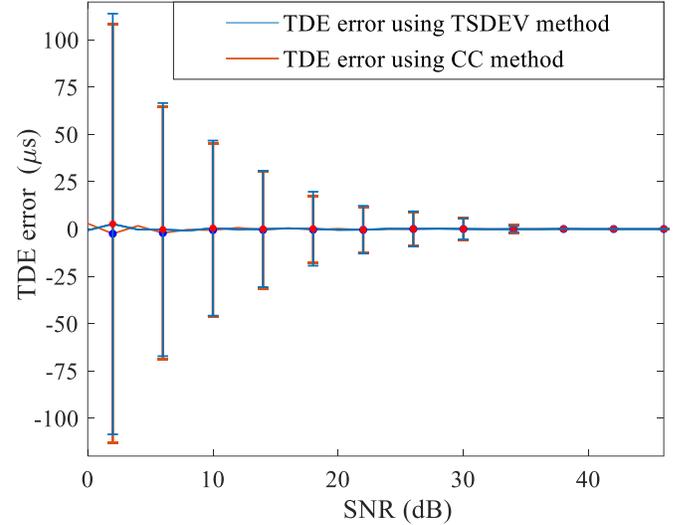

Fig. 1. TDE error using TSDEV method and cross-correlation method when the SNR is changed from 45 dB to 0 dB. The phase changing signals used are sine wave with 20 Hz frequency and $100\mu s$ time delay $\tau_0$. Intercepted length is four periods of the sine wave (0.2 s). CC: cross-correlation.

The detected signals can be modeled as:

$$x_1(t) = s(t) + n_{w1}(t), \quad (7)$$
$$x_2(t) = s(t-\tau_0) + n_{w2}(t). \quad (8)$$

Here, $n_{w1}(t)$ and $n_{w2}(t)$ are the uncorrelated white noise. Normally, the fluctuations of $R_{nw}(\tau)$ (cross-correlation function between $n_{w1}(t)$ and $n_{w2}(t)$) and $TSDEV_{nw}^2(\tau)$ (TSDEV value between $n_{w1}(t)$ and $n_{w2}(t)$) are both small as $\tau$ changes, they hardly affect TDE result at high SNR. With the decrease of SNR, it will add random error to the TDE result.

We used sinusoidal signals to simulate these two methods' TDE results. The frequency of the sine wave is 20 Hz. Time delay $\tau_0$ is $100\mu s$. Intercepted length is four periods of the sine wave (0.2 s). Gaussian white noise with different power is added, which worsens the SNR from 45 dB to 0 dB. Using two TDE methods to determine the time delay $\tau_0$, the standard deviation of TDE error at different SNRs is shown in Fig.1. It shows that the results of cross-correlation and TSDEV methods is the same for the condition of white noise.

## B. Time delay estimation at arbitrary intercepted length

In actual case, the measurement time and intercepted length of signal are always limited or random. In this part, we simulate the signal with arbitrary intercepted length and analyze the accuracy of two TDE methods.

Based on (4), we can calculate the cross-correlation function of phase changing signals, and that of noise separately. For the convenience of calculation, we still take the sinusoidal signal as an example. The cross-correlation function of phase changing signals can be written as:

$$R_s(\tau) = \frac{1}{T_w}\int_{t_0}^{t_0+T_w}\left\{\left[\sin(\omega t)\right]\left[\sin\left[\omega(t-\tau_0+\tau)\right]\right]\right\}dt$$
$$= \frac{1}{2}\cos[\omega(\tau-\tau_0)] - \frac{1}{2\omega T_w}\sin\omega T_w \cdot \cos[2\omega t_0 + \omega T_w + \omega(\tau-\tau_0)]. \quad (9)$$

Here, $t_0$ is initial phase of the intercepted signal segment, $T_w$ is length of the intercepted window. If the white noise is considered, $R(\tau) = R_s(\tau) + R_{nw}(\tau)$, which will just add random error to the TDE result. When the length of window $T_w$ is arbitrary, in addition to the first term $\frac{1}{2}\cos[\omega(\tau-\tau_0)]$, second term $\frac{1}{2\omega T_w}\sin\omega T_w \cdot \cos[2\omega t_0 + \omega T_w + \omega(\tau-\tau_0)]$ will appear. Therefore, the overall cross-correlation result will not always reach maximum value at $\tau=\tau_0$, and will be affected by parameters of $t_0$ and $T_w$. Because the initial phase $t_0$ is uncertain, only if $T_w = N\cdot\frac{T}{2}$ occurs (N is the positive integer), will the second item be 0. It means that only when the intercepted length is infinite or integer multiples of half period, can cross-correlation method obtain the precise TDE result.

For the TSDEV method, the TSDEV value of signals with arbitrary intercepted length is given by:

$$TSDEV_s^2(\tau) = \frac{1}{T_w}\int_{t_0}^{t_0+T_w}\{\sin(\omega t) - \sin[\omega(t-\tau_0+\tau)] - C(\tau)\}^2 dt$$
$$= A(\tau)\cdot\sin^2\left[\frac{1}{2}\omega(\tau-\tau_0)\right] + B(\tau)\cdot\sin\left[\frac{1}{2}\omega(\tau-\tau_0)\right] + C^2(\tau), \quad (10)$$

where

$$A(\tau) = \frac{1}{\omega T}\{2\omega T + \sin[\omega(2t_w + 2T + \tau - \tau_0)] - \sin[\omega(2t_w + \tau - \tau_0)]\},$$

$$B(\tau) = \frac{4C(\tau)}{\omega T}\left\{\sin\left[\omega(t_w + T + \frac{1}{2}\tau - \frac{1}{2}\tau_0)\right] - \sin\left[\omega(t_w + \frac{1}{2}\tau - \frac{1}{2}\tau_0)\right]\right\},$$

and $C(\tau) = \frac{1}{T}\int_{t_w}^{t_w+T}\{\sin(\omega t) - \sin[\omega(t-\tau_0+\tau)]\}dt$. When considering white noise, $TSDEV^2(\tau) = TSDEV_s^2(\tau) + TSDEV_{nw}^2(\tau)$, which will just add random error to the TDE result. Although (10) has 3 terms, it will not influence the TDE result. All these terms will reach 0 at $\tau=\tau_0$, and the overall TSDEV value will reach minimum value at $\tau=\tau_0$. Considering that arbitrary signal can be decomposed into multiple signals with single frequency by FS, this conclusion is still reliable for arbitrary signal.

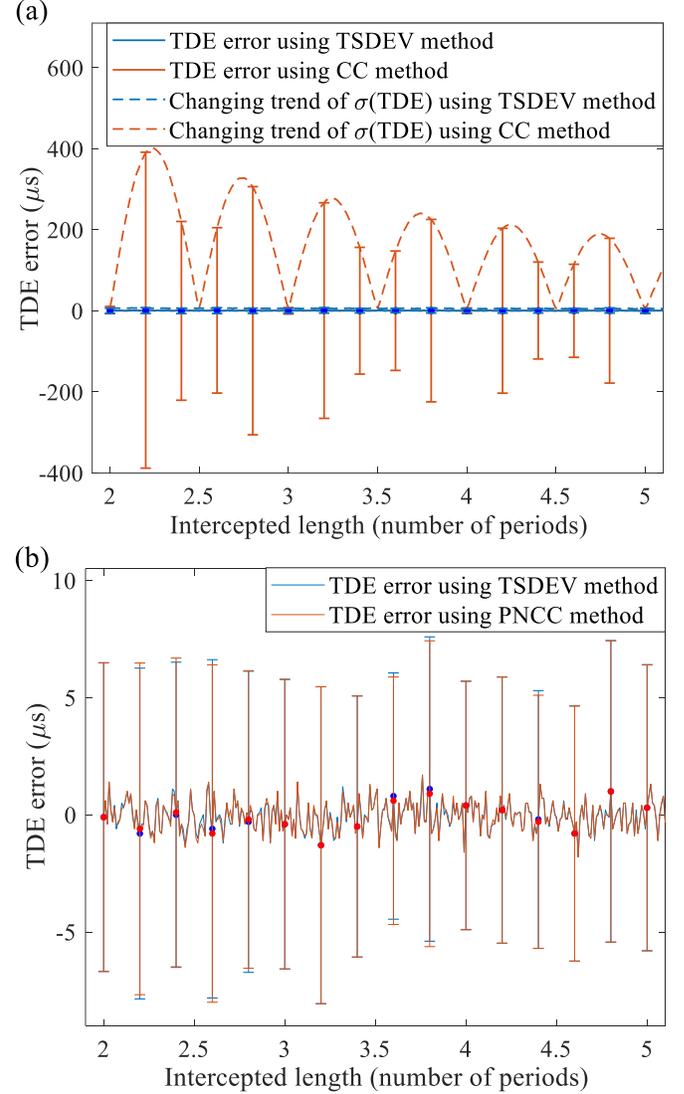

Fig. 2. The TDE errors under different intercepted length. CC: cross-correlation, PNCC: power normalized cross-correlation. The phase changing signal used is 20 Hz sine wave, time delay $\tau_0 = 100\mu s$ and SNR=30dB. (a) The comparison between TSDEV and CC methods. Red error bar: TDE errors using correlation method. Blue error bar: TDE errors using TSDEV method. Red dash lines: Changing trend of $\sigma(TDE)$ using cross-correlation method. Blue dash lines: Changing trend of $\sigma(TDE)$ using TSDEV method. (b) The comparison between TSDEV and PNCC methods. Red: TDE errors using PNCC method. Blue: TDE errors using TSDEV method.

In order to verify the conclusion above, we simulate the TDE results of these two methods using signals with different intercepted length. Parameters of two signals are the same as before (20 Hz frequency, $\tau_0 = 100\mu s$, and SNR=30 dB). According to theoretical analysis, the result of cross-correlation is affected by intercepted length $T_w$ and initial phase $t_0$. For each intercepted length $T_w$, we change initial phase $t_0$, and will obtain a set of TDE values. The standard deviation of TDE error $\sigma(TDE)$ and its changing trend with intercepted length $T_w$ are shown in Fig. 2(a). It can be found that cross-correlation method can only get precise results when the intercepted length is integer multiples of half period $T_w = N \cdot \frac{T}{2}$. However, for TSDEV method, the TDE results will not be affected by intercepted length and initial phase, and its standard deviation is always kept at a small value, which is just induced by the white noise.

The reason for this phenomenon is that, cross-correlation can be regarded as the inverse Fourier transform of cross-power spectrum between two intercepted signals. Its result will be affected by two segments' consistency and average power. Average power of a signal $x(t)$ is defined by [29]:

$$P_x = \lim_{T_w \to \infty} \frac{1}{T_w} \int_{-T_w/2}^{T_w/2} |x(t)|^2 dt. \quad (11)$$

For an intercepted sinusoidal signal segment, the length of segment is from $t_0$ to $t_0 + T_w$, and its average power can be written as:

$$P_x = \lim_{T_w \to \infty} \frac{1}{T_w} \int_{t_0}^{t_0+T_w} |\sin(\omega \cdot t)|^2 dt$$
$$= \frac{1}{2} + \frac{1}{4\omega \cdot T_w} \{\sin[2\omega \cdot (t_0 + T_w)] - \sin(\omega \cdot t_0)\}. \quad (12)$$

It shows that average power is constant only if $T_w = N \cdot \frac{T}{2}$, which is consistent with the condition for obtaining precise TDE result using cross-correlation method in (9). It indicates that for signal segments with constant power, the cross-correlation method can work out the real time delay. However, if segment's power changes during time shifting, the position of correlation peak will deviate from the real one. Thus the TDE result of cross-correlation may be erroneous.

To solve this problem, an improved cross-correlation method may be used, in which the average power is continuously normalized for each time shifting step. We can get the power normalized cross-correlation (PNCC) function $R_{PNCC}(\tau)$:

$$R_{PNCC}(\tau) = \frac{\frac{1}{T_w}\int_{t_0}^{t_0+T_w} x_1(t)x_2(t+\tau)dt}{\sqrt{\frac{1}{T_w}\int_{t_0}^{t_0+T_w} x_1^2(t)dt \cdot \frac{1}{T_w}\int_{t_0}^{t_0+T_w} x_2^2(t+\tau)dt}}$$
$$= \frac{R(\tau)}{\sqrt{P_1 \cdot P_2(\tau)}}. \quad (13)$$

Fig. 2(b) shows the standard deviation of TDE error with arbitrary intercepted signal segments, for TSDEV and PNCC methods respectively. They have the same accuracy. However, PNCC is different from the normalized cross-correlation function [20], since the power compensation happens in every time shifting step. Consequently, the calculation speed of PNCC method is much slower than TSDEV method.

Different from cross-correlation, the formula of TSDEV method (5) can be further expanded as,

$$TSDEV^2(\tau) = \frac{1}{T_w}\int_{t_0}^{t_0+T_w}\left[x_1(t) - x_2(t+\tau) - C(\tau)\right]^2 dt$$
$$= \frac{1}{T_w}\int_{t_0}^{t_0+T_w} x_1^2(t)dt + \frac{1}{T_w}\int_{t_0}^{t_0+T_w} x_2^2(t+\tau)dt$$
$$- \frac{1}{T_w}\int_{t_0}^{t_0+T_w} 2x_1(t)x_2(t+\tau)dt - C^2(\tau) \quad (14)$$
$$\geq 0.$$

Based on the Cauchy inequality, $TSDEV^2(\tau) \geq 0$ is always true, only if $x_1(t) = x_2(t+\tau)$, $TSDEV^2(\tau) = 0$ occurs. Thus, the TSDEV method can obtain precise TDE results at arbitrary intercepted length.

*C. The case of common noise*

Beside the uncorrelated white noise, there will be common noise $n_c(t)$ in two detected signals induced by measurement system and environment. In this part we analyze the influence of common noise, which can be modeled as:

$$x_1(t) = s(t) + n_c(t), \quad (15)$$
$$x_2(t) = s(t-\tau_0) + n_c(t). \quad (16)$$

Here, $n_c(t)$ stands for the common noise, including linear drift, low-frequency environmental noise and wideband noise caused by measurement system, etc.

For linear drift $n_c(t) = k \cdot t$, the influence on the result of TSDEV value is:

$$TSDEV_{nc}^2(\tau) = \frac{1}{T_w}\int_{t_0}^{t_0+T_w}\left[k \cdot t - k \cdot (t+\tau) - C(\tau)\right]^2 dt$$
$$= \frac{1}{T_w}\int_{t_0}^{t_0+T_w}\left[-k \cdot \tau + k \cdot \tau\right]^2 dt \quad (17)$$
$$= 0.$$

Here, $C(\tau) = \frac{1}{T_w}\int_{t_0}^{t_0+T_w}[k \cdot t - k \cdot (t+\tau)]dt = -k \cdot \tau$. No matter how large k is, $TSDEV_{nc}^2(\tau)$ (TSDEV value of $n_c(t)$) is always 0. On the contrary, cross-correlation method cannot handle nonstationary signals such as linear drift, because correlation between drifts will change with time shifting.

During the simulation, two sinusoidal signals are the same as before (20 Hz frequency, $\tau_0 = 100\mu s$, and SNR is 30 dB). For different drift slopes, TDE errors of PNCC and TSDEV methods are shown in Fig.3(a). For PNCC method, the standard deviation of TDE errors at different slopes have no changes, but the center value is shifted when the slope increases. While for the TSDEV method, the mean value is always precise, and TDE errors are not affected by the linear drift.

For low-frequency environmental noise $n_c(t) = \sin(\omega t)$, traditionally, it can be suppressed by cleverly designing the detection system structure to achieve a differential operation [12,30]. Another method is to reduce the low-frequency noise using filters [31-32]. However, these two methods will reduce the SNR or make the vibration induced phase changing signal damaged, which leads to the loss of precision. For TSDEV method, it uses the principle of differential, which can suppress the influence of low-frequency noise.

Two sinusoidal signals are used to simulate, which are the same as before (20 Hz frequency, $\tau_0 = 100\mu s$, and SNR is 30 dB). Considering the common noise whose frequency changes from 0.01 Hz to 0.5 Hz, the results are shown in Fig. 3(b). For PNCC method, When the frequency is higher than 0.2 Hz, TDE error will deviate from true value seriously. Although the results of TSDEV is also shifted as the frequency increases, the degree is relatively small.

For higher frequency common noise (normally, $n_c(t)$ is a wideband noise), the effect of differential operation is limited, the results of TSDEV will also deviated from true value, which is shown as follows:

$$TSDEV^2(\tau)$$
$$= \frac{1}{T_w} \int_{t_0}^{t_0+T_w} \left\{ \left[s(t) + n_c(t)\right] - \left[s(t-\tau_0+\tau) + n_c(t+\tau)\right] - C(\tau) \right\}^2 dt$$
$$= TSDEV_s^2(\tau) + TSDEV_{nc}^2(\tau).$$
(18)

Here, $TSDEV_{nc}^2(\tau)$ represents the TSDEV of $n_c(t)$. $TSDEV_s^2(\tau)$ will get the minimum value at $\tau = \tau_0$, but $TSDEV_{nc}^2(\tau)$ will get the minimum value at $\tau = 0$. Therefore, the result of TDE will shift to the direction of zero time delay (the TDE error will shift to the negative value). The same problem exists in cross-correlation method.

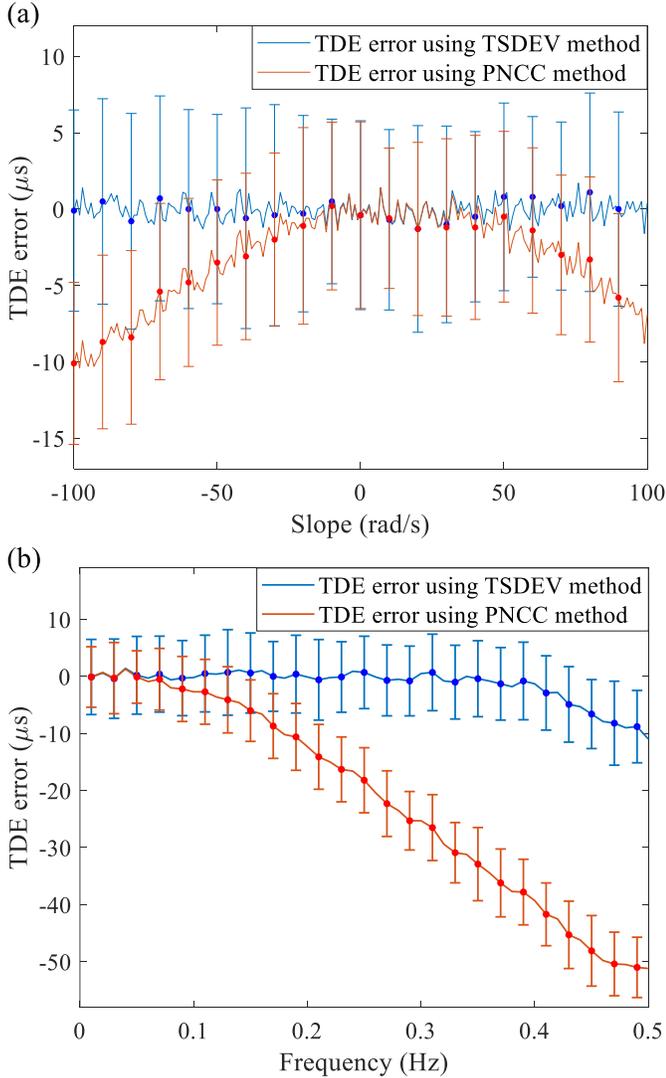

Fig. 3. (a) The simulated TDE errors when linear drift with different slopes (from -100 rad/s to 100 rad/s) is added to the phase changing signals. Red: TDE errors using PNCC method. Blue: TDE errors using TSDEV method. (b) The simulated TDE errors when low-frequency noise with different frequency (from 0.01 Hz to 0.5 Hz) is added to the phase changing signals. Red: TDE errors using PNCC method. Blue: TDE errors using TSDEV method. The phase changing signal used is 20 Hz sine wave, time delay $\tau_0 = 100\mu s$ and SNR=30dB. Intercepted length is four periods of the sine wave (0.2 s).

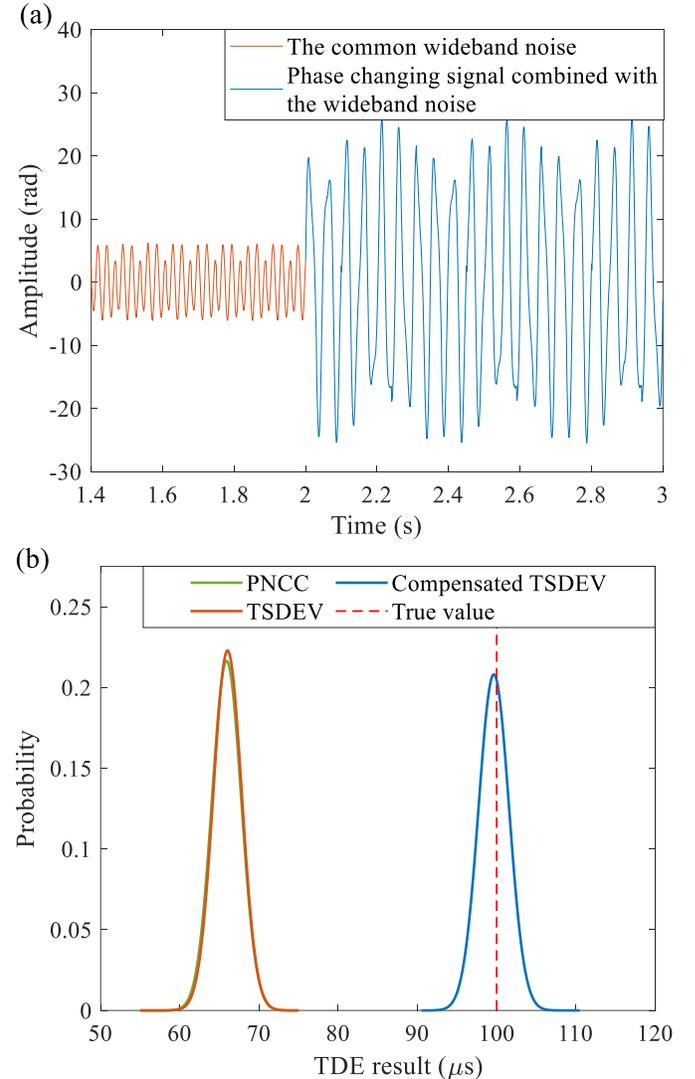

Fig. 4. (a) The simulated phase changing signal with wideband common noise. Red part: noise segment centered at 40 Hz with 20 Hz bandwidth. Blue part: phase changing signal combined with the wideband common noise (The phase changing signal used is 20 Hz sine wave, time delay $\tau_0 = 100\mu s$ and SNR=30 dB). (b) TDE results using PNCC, TSDEV and compensated TSDEV methods. Red dash line: the true time delay value ($100\mu s$). Green line: TDE results'

distribution using PNCC method. Red line: TDE results' distribution using TSDEV method. Blue line: TDE results' distribution using compensated TSDEV method.

In order to solve this problem, we propose a compensated TSDEV method. For a stable system, the statistical characteristics of its noise are unchanged in a short time. If we select a signal segment with only system noise whose duration is close to that of detected signal segments, their TSDEV can be used to represent $TSDEV_{nc}^2(\tau)$. We can use it to compensate and correct the results of TSDEV. It is mainly because TSDEV value of signals and TSDEV value of noise can be easily separated. However, for PNCC method, its numerator and denominator are both affected by the noise, its value isn't the superposition of signals' PNCC function and noise's PNCC function ( $PNCC(\tau) \neq PNCC_s(\tau) + PNCC_n(\tau)$ ). Thus, the same simple compensation scheme can't be used in PNCC method.

We use a signal centered at 40 Hz with 20 Hz bandwidth to simulate the wide band common noise, and the target vibration induced signal is the same as before (20 Hz frequency, $\tau_0 = 100\mu s$, and signal-to-white noise ratio is 30 dB). In Fig. 4(a), the red line shows the wideband common noise, and the blue line shows the superposition of the wideband common noise with the target vibration induced signal. The distribution of TDE results is shown in Fig. 4(b). The TDE results using PNCC or TSDEV methods are shifted from the true value ( $100\mu s$ ), and it becomes correct using the compensated TSDEV method. Furthermore, comparing three Gaussian curves in Fig. 4(b), compensation scheme will only change the mean of TDE results, their standard deviation will not be affected. Therefore, compensated TSDEV method still have high precision the same as TSDEV before compensation.

IV. EXPERIMENTAL SYSTEM AND LOCALIZATION PRINCIPLE

We set up a laser interferometer based on forward transmission system to detect vibration along the fiber link, which is shown in Fig. 5. In our sensing system, the laser light is divided into two beams by an 80/20 coupler, which enter the sensing route and reference route, respectively. The light in the sensing route is frequency shifted by acousto-optic modulator (AOM), and divided into two different directions by a 50/50 coupler. One transmits clockwise (CW beam) and another transmits counter clockwise (CCW beam) along two fiber spools with the length of 50 km (49.49 km) and 10 km (9.84 km). Between these two spools, a fiber stretcher (FST) is used to generate vibration. Then, two beams are separated by the circulators, and finally interfere with the reference beam at two photodetectors (PD), respectively. Through analyzing the phase changes of two detected signals, the vibration monitoring and positioning can be realized [33].

When vibration occurs, corresponding phase variations will be detected on PD1 and PD2, and the time delay between these two signals can be used to determine the location of vibration event.

$$x_{CW}(t) = s\left(t - \frac{L-l}{c/n}\right) + n_1(t), \quad (19)$$

$$x_{CCW}(t) = s\left(t - \frac{l}{c/n}\right) + n_2(t). \quad (20)$$

Here, $x_{CW}(t)$ and $x_{CCW}(t)$ represent two beams' phase changing respectively. $l$ refers to the distance between OC2 and vibration source (FST) clockwise. $L$ is the length of fiber link, c is the speed of light and $n$ is the refractive index of optical fiber. $n_1(t)$ and $n_2(t)$ is noise, including uncorrelated white noise and common noise. By estimating the real time delay $\tau_0$ between $s\left(t - \frac{l}{c/n}\right)$ and $s\left(t - \frac{L-l}{c/n}\right)$, we can locate the vibration point via:

$$l = \frac{1}{2}\left(L - \frac{c}{n}\tau_0\right). \quad (21)$$

It shows that, in the optical fiber vibration monitoring system, the vibration positioning accuracy mainly depends on the TDE results.

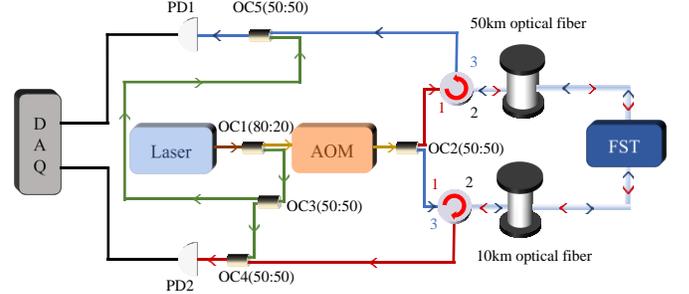

Fig. 5. The experimental set up of laser interferometer based on forward transmission scheme. Laser: ultra-narrow linewidth laser, PD: photodetector, AOM: acoustic optical modulator, OC: optical coupler, FST: fiber stretcher, DAQ: data acquisition. Red part: CW beam. Blue part: CCW beam. Green part: reference beams.

V. EXPERIMENTAL RESULT

The vibration signal we used is a wide band signal centered at 150Hz with the bandwidth of 100 Hz. The detected phase changing signals are shown in Fig. 6(a). We can see there is a slow fluctuation, which is the superposition of linear drift and low-frequency environmental noise. In the nonvibration signal segment, two detected noise are mainly common noise. And in the vibration signal segment, there is an obvious time delay between two signals (from the enlarged plots). These are all consistent with our theoretical analysis.

The time window of 0.126 s is used to intercept signals. The localization results of several methods are shown in Fig. 6(b). Because the initial phase we chose is random and intercepted window's length isn't integer multiples of one period, TDE results of cross-correlation method has a large standard deviation, which reaches ~208.4 m. On the contrast, the results' standard deviation of TSDEV method is only 22.3 m. If PNCC method is used, similar precision as TSDEV method can be realized. However, the average of localization results obtained by TSDEV and PNCC methods both deviate from the true value

~67.8 m. All the results shift to the direction of entire link's center, whose time delay is zero. The compensated TSDEV method successfully eliminates this offset, and the localization precision doesn't lose.

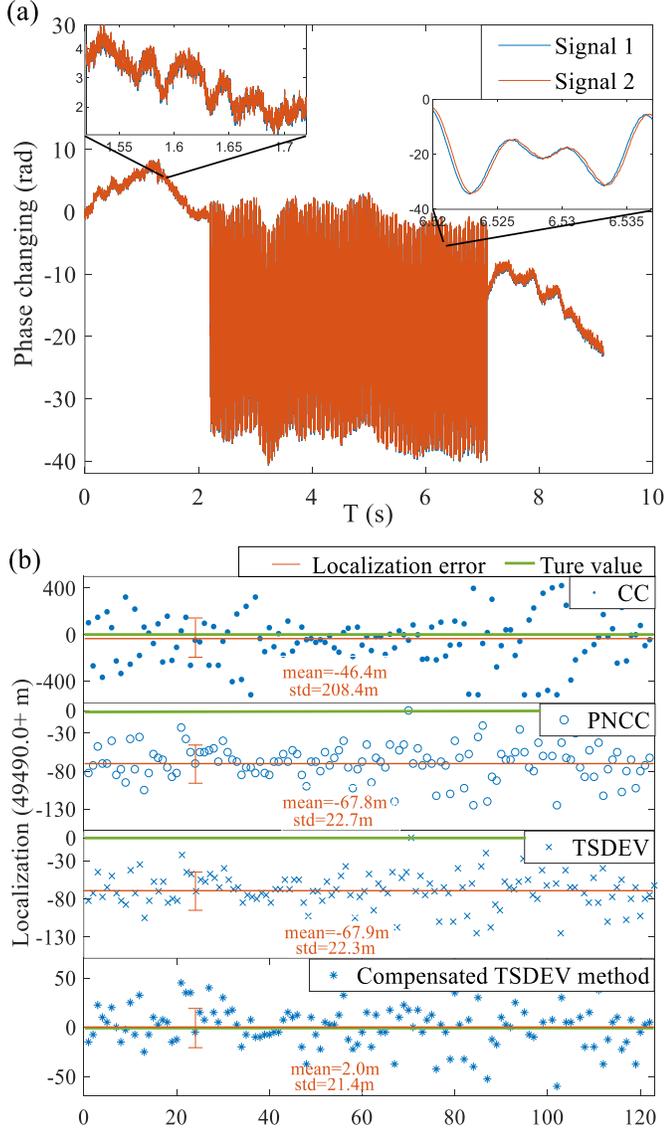

Fig. 6. (a) The detected phase changing signals with background noise. The enlarged plot in the left corner shows the common noise without time delay. The enlarged plot in the right corner shows the vibration signals with obvious time delay. The vibration is a wideband signal centered at 150 Hz with the bandwidth of 100 Hz. (b) Localization results using CC, PNCC, TSDEV and compensated TSDEV methods. Green line: true location of the vibration, 49.4900km apart from OC2 clockwise. Red line: estimated location with different method. CC: cross-correlation, mean: the error from the true location, std: standard deviation.

In this experiment, in order to compare the advantages and disadvantages of several methods obviously, we use a low amplitude vibration whose phase change is only 35 rad. More precise localization results of TSDEV are shown in [27], in which the standard deviation of localization results 2.5 m is realized and a commercial buried fiber cable is used to detect traffic information and localize vibration in Beijing.

## VI. CONCLUSION

We propose a novel TSDEV method. Compared with cross-correlation, the proposed TSDEV method can be used at arbitrary intercepted length. In addition, because TSDEV method contains the principle of difference, it has a good suppression on linear drift and low-frequency noise. For common noise with wide band, a compensation scheme is applied, and simulation verifies its effectiveness.

We demonstrate a laser interferometer based on forward transmission scheme to detect vibration along the fiber link. For a wideband signal centered at 150 Hz with the bandwidth of 100 Hz. The typical localization error is 2 m with standard deviation of 21.4 m using TSDEV method. It stands in clear contrast to the results of cross-correlation method, whose localization error is 46.4 m and standard deviation is 208.4 m. The experimental results show that compared with cross-correlation method, TSDEV has the same resistance to white noise, but has fewer boundary conditions and better suppression on temperature drift or common noise, which leads to more precise TDE results.